\documentclass[aps,draft]{revtex4}
\usepackage[dvips]{graphicx}
\begin{document}

\title{Elementary Excitations in Quantum Fermi Liquid}
\author{ Nodar L.Tsintsadze }
\author{ Levan N.Tsintsadze }
\thanks{Graduate School of Science, Hiroshima University, Higashi-Hiroshima, Japan}
\affiliation{Department of Plasma Physics, E.Andronikashvili
Institute of Physics, Tbilisi, Georgia}

\date{\today}

\begin{abstract}
Landau's theory of Fermi liquids is generalized by incorporating the de Broglie waves diffraction.
A newly derived kinetic equation of the Fermi particles is used to derive a general dispersion relation and the excitation of zero sound is studied. A new mode is found due to the quantum correction. It is shown that the zero sound can exist even in an ideal Fermi gas. We also disclose a new branch of frequency spectrum due to the weak interaction.
\end{abstract}


\maketitle

It is well known that at temperatures $\sim 1-2^oK$ only two quantum liquids exist in nature, the isotopes of helium $H_e^3$ and $H_e^4$, and all other substances solidify. The peculiarly weak interaction between the helium atoms is the reason for helium to remain liquid. Based on this fact, namely that in $H_e^3$ the weak interactions take place between atoms at sufficiently low temperatures, Landau has created the theory of Fermi liquid (FL) \cite{lan}. In which he took into account only the weakly excited energy levels of the liquid, lying fairly close to the ground state. Landau assumed that any weakly excited state of a macroscopic body can be represented as an assembly of separate elementary excitations (quasiparticles). Moreover, the elementary excitations are represented as the collective motion of atoms in a liquid and it cannot be identified with individual atoms. Therefore, an important characteristic of the energy spectrum is the establishment of the dispersion relation $\varepsilon(p)$ for elementary excitations.
Landau has then shown that the undamped zero sound can exist in an almost ideal Fermi gas. A weak electrostatic interactions of electrons with each other were discussed and spectrum of elementary excitations was investigated in Refs.\cite{gol}-\cite{bp}. The perturbation theory was used in Ref.\cite{ks}, and then in Ref.\cite{sil} to study oscillations of the uncharged Fermi gas.

In this Letter, we extend Landau's theory of Fermi liquids by taking into account the de Broglie waves diffraction, and show that even in an ideal Fermi gas, when the interaction between atoms absent, the dispersion relation of the zero sound preserves the form. We also disclose a new branch of frequency spectrum due to the weak interaction.

To consider the problem of existence of the zero sound in $H_e^3$ liquid, we employ a novel quantum kinetic equation derived by us in Ref.\cite{tsin}. We note here that the difference between the Landau kinetic equation and ours is that in our equation an additional term, namely the Madelung term is incorporated due to the diffraction of de Broglie waves.

Non-equilibrium states of a Fermi quantum liquid are described by the one particle
distribution function $f(\vec{r},\vec{p},t,\vec{\sigma} ),$ which satisfies the
quantum Boltzmann equation \cite{tsin}
\begin{eqnarray}
\label{qb}
\frac{\partial f}{\partial t}+( \vec{v}\cdot\nabla )f+\frac{d\vec{p}}{dt}\ \frac{\partial
f}{\partial \vec{p}}=C(f) \ ,
\end{eqnarray}
where $\frac{d\vec{r}}{dt}=\vec{v}=\frac{\partial \varepsilon}{\partial \vec{p}}\ ,\ $$\ \frac{d\vec{p}}{dt}=-\frac{\partial \varepsilon}{\partial \vec{r}}+\frac{\hbar^2}{2m}\nabla \frac{1}{\sqrt{
n}}\Delta \sqrt{n }\ ,\ $ $\varepsilon$ is the energy of the quasiparticle and in general its a matrix with respect to the spin variables $\vec{\sigma}$, m and n are mass and density of particles, respectively, $\hbar$ is the Planck constant divided by $2\pi$, and $C(f)$ is the collision integral, which describes the variation of the distribution function due to particle collisions.

Note that when the spin of particles is taken into account, the distribution function is an operator with respect to the spin variables $\sigma $. The quasiparticles in a Fermi liquid have spin  1/2. However, there is a wide range of problems in which it is sufficient to consider a distribution independent of spin variables, so that $f$ becomes the ordinary quasiclassical distribution function $f(\vec{r},\vec{p},t)$. We recall here that the condition for quasiclassical motion is that the de Broglie wavelength $\lambda_d=\hbar/p_F$ ($p_F$ is the Fermi momentum) of the particle must be very small compared with the characteristic length L, over which $f(\vec{r},\vec{p},t)$ varies considerably.

Following the Landau's theory, hereafter, we consider FL as a spinless, and the energy $\varepsilon$ of quasiparticles is a functional of the distribution function; a variation of distribution function
\begin{eqnarray}
\label{perd}
f(\vec{r},\vec{p},t)=f_0(\vec{p})+\delta f(\vec{r},\vec{p},t)
\end{eqnarray}
produces a variation of energy given by
\begin{eqnarray}
\label{enc}
\delta\varepsilon=\int\varphi(\vec{p},\vec{p}\prime)\delta f(\vec{r},\vec{p},t)\ \frac{2d^3p}{(2\pi\hbar)^3}\ ,
\end{eqnarray}
where the factor 2 appears due to a spin, $\ f_0(\vec{p})$ and $\varphi(\vec{p},\vec{p}\prime)$ are the equilibrium distribution function and the quasiparticle interaction function, respectively; in a Fermi gas $\varphi=0$. Thus the distribution function (\ref{perd}) refers to the energy of quasiparticle
\begin{eqnarray}
\label{eqp}
\varepsilon=\varepsilon_0(\vec{p})+\delta\varepsilon(\vec{r},\vec{p},t)\ ,
\end{eqnarray}
where $\varepsilon_0(\vec{p})$ is the energy corresponding to the equilibrium state and has a certain physical significance near the surface of the Fermi sphere, i.e. only quasiparticles with momenta p, such that $\mid p-p_F\mid\ll p_F$, have any real meaning. We can therefore expand $\varepsilon_0(p)$ in powers of the difference $p-p_F$ to obtain
\begin{eqnarray}
\label{peq}
\varepsilon_0=\varepsilon_F+\Bigl(\frac{\partial\varepsilon_F}{\partial p}\Bigr)_{p=p_F}\ (p-p_F) \ ,
\end{eqnarray}
where $\Bigl(\frac{\partial\varepsilon_F}{\partial p}\Bigr)_{p=p_F}=v_F$ is the speed of quasiparticles at the Fermi surface.

Now Eq.(\ref{eqp}) reads
\begin{eqnarray}
\label{nr}
\varepsilon=\varepsilon_F+v_F(p-p_F)+\int\varphi(\vec{p},\vec{p}\prime)\delta f(\vec{r},\vec{p}\prime,t)\ \frac{2d^3p\prime}{(2\pi\hbar)^3}\ .
\end{eqnarray}
Near the surface of the Fermi sphere the variation of distribution function  $\delta f(\vec{r},\vec{p}\prime,t)$ is appreciably different from zero, i.e. the magnitude $p\prime=p=p_F$. The same is true for the function $\varphi(\vec{p},\vec{p}\prime)$. So that both depend only on directions of the vectors $\vec{p}$ and $\vec{p}\prime$. Hence, the quasiparticle interaction function $\varphi$ and $\delta f$ can be expressed at the Fermi surface as
\begin{eqnarray}
\label{pfs}
\varphi(\vec{p},\vec{p}\prime)=\frac{\pi^2\hbar^3}{m^*p_F}\ Q(\theta)
\end{eqnarray}
\begin{eqnarray}
\label{dffs}
\delta f=\delta(\varepsilon-\varepsilon_F)F(\vec{n}\prime,\vec{r},t) \ ,
\end{eqnarray}
where  $m^*=p_F/v_F$ is the effective mass of quasiparticle, $\vec{n}\prime$ is the unit vector in the direction of $\vec{p}\prime$, and $Q(\theta)$ is the function of the angle $\theta$ between $\vec{p}$ and $\vec{p}\prime$.

We now employ the equation (\ref{qb}) to study the propagation of small perturbations in the Fermi quantum liquid. We substitute the expressions (\ref{perd}) and (\ref{nr}) into Eq.(\ref{qb}) and linearize it with respect to the perturbation $\delta f$ to obtain
\begin{eqnarray}
\label{lin}
\frac{\partial\delta f}{\partial t}+( \vec{v}\cdot\nabla )\delta f-\nabla\delta\varepsilon\ \frac{\partial
f_0}{\partial \vec{p}}+\frac{\hbar^2}{4m}\nabla\cdot\Delta\frac{\delta n}{n_0}\ \frac{\partial
f_0}{\partial \vec{p}}=C(\delta f) \ .
\end{eqnarray}
We look for wave solutions in space and time for $F(\vec{n}\prime,\vec{r},t)$, assuming that it is proportional to $\exp{i(\vec{k}\vec{r}-\omega t)}$.

Taking into account Eqs.(\ref{nr})-(\ref{dffs}), $\nabla\varepsilon$ can be written as
\begin{eqnarray}
\label{nec}
\nabla\delta\varepsilon=\int\frac{d\Omega\prime}{2}Q(\theta\prime)\ \nabla F(\vec{n}\prime,\vec{r},t)=\iota\vec{k}\int\frac{d\Omega\prime}{2}\ Q\ F(\vec{n}\prime) \ ,
\end{eqnarray}
where $d\Omega=sin\theta d\theta$.

Whereas the density variation reads
\begin{eqnarray}
\label{dver}
\frac{\delta n}{n_0}=\frac{3m}{p_F^2}e^{\iota(\vec{k}\vec{r}-\omega t)}\int\frac{d\Omega\prime}{2}\ F(\vec{n}\prime)
\end{eqnarray}
and
\begin{eqnarray}
\label{pfo}
\frac{\partial f_0}{\partial \vec{p}}=-\vec{v}\delta(\varepsilon-\varepsilon_F)\ .
\end{eqnarray}
The wave propagation in the Fermi liquid, as shown by Landau, is possible if $\omega\tau\gg 1$ ($\tau$ is the mean free time), which means that collisions of quasiparticles are unimportant and the collision integral can be neglected.

Use of Eqs.(\ref{nec})-(\ref{pfo}) in Eq.(\ref{lin}) then yields the relation
\begin{eqnarray}
\label{yie}
F(\vec{n})=\frac{kv_Fcos\theta}{\omega-kv_Fcos\theta}\ \left\{\int\frac{d\Omega\prime}{2}Q(\theta\prime)F(\vec{n}\prime)+\frac{3\hbar^2k^2}{4p_F^2}
\int\frac{d\Omega\prime}{2}F(\theta\prime)\right\}\ .
\end{eqnarray}

If we suppose that $Q(\theta\prime)$ is constant, i.e. $Q(\theta\prime)=Q_0$, then from Eq.(\ref{yie}) after the integration with respect to the angle, we obtain
\begin{eqnarray}
\label{disp}
1-\Bigl(Q_0+\frac{3\hbar^2k^2}{4p_F^2}\Bigr)\ \frac{1}{2}\int_0^\pi\ \frac{d\theta\ sin\theta\ cos\theta}{S-cos\theta}=0 \ ,
\end{eqnarray}
where $S=\frac{\omega}{kv_F}$.

Integrating Eq.(\ref{disp}), we finally arrive at the dispersion relation
\begin{eqnarray}
\label{dispr}
1+\Bigl(Q_0+\frac{3\hbar^2k^3}{4p_F^2}\Bigr)\ \Bigl(1-\frac{\omega}{2kv_F}\ln\frac{\omega+kv_F}{\omega-kv_F}\Bigr)=0 \ ,
\end{eqnarray}
solution of which describes undamped waves.

In the case when $\omega=kv_F+\gamma\ $, $\ \mid\gamma\mid\ll kv_F,\ $ Eq.(\ref{dispr}) admits the solution 
\begin{eqnarray}
\label{ads}
\omega =kv_F\Bigl(1+2e^{-2\{1+
\frac{1}{Q_0+(3/4)(k\lambda_d)^2}\}}\Bigr)\ .
\end{eqnarray}
Thus we have generalized the Landau's dispersion relation of zero sound including the quantum correction. The condition of the existence of zero sound is that $Q_0$ and $(k\lambda_d)^2$ must be less than one. The latter is always much less than one. Note that in Landau's theory $Q_0$ must also be less than one.

It should be emphasized that in the absence of interaction between atoms, i.e. $Q_0=0$, as follows from Eq.(\ref{ads}), the undamped zero sound waves can be excited in an ideal Fermi gas due to the de Broglie waves diffraction, the spectrum of which is
\begin{eqnarray}
\label{uzs}
\omega =kv_F\Bigl(1+2e^{-2\{1+
\frac{1}{(3/4)(k\lambda_d)^2}\}}\Bigr)\ .
\end{eqnarray}
It should be noted that the dispersion relation is purely quantum.

The definition of $Q_0$ (relation (\ref{pfs})) admits $Q_0$ to be more than one. In this case Eq.(\ref{dispr}) has another interesting solution. Namely, for the case $\omega\gg kv_F$ from Eq.(\ref{dispr}) follows the dispersion equation
\begin{eqnarray}
\label{adis}
\omega^2=\frac{Q_0}{3}k^2v_F^2+\omega_q^2 \ ,
\end{eqnarray}
where $\omega_q=\frac{\hbar k^2}{2m}$ is the quantum frequency.
If there are no interactions between atoms, i.e. $Q_0=0$, we recover the frequency of quantum oscillations of free atoms. We specifically note here that the frequency (\ref{adis}) is novel and is the product of weak interactions of atoms.

In the above consideration we assumed that the temperature $T\ll\hbar\omega\ll\varepsilon_F$, which means that the width of transitional zone of the distribution function is $\hbar\omega$. Taking into account the collision between quasiparticles, it is evident that the waves (\ref{ads})-(\ref{adis}) will be weakly damped in the case when $\omega\tau\gg 1$. As mentioned above, the quasiparticle may be regarded as a particle in the self-consistent field of surrounding particles. The collision of two quasiparticles in such field is accompanied by a change of their total energy and momentum by $\hbar\omega$ and $\hbar\vec{k}$, respectively. In the process, a sound wave is absorbed or emitted in collisions. The overall effect of such collisions is to reduce the total number of sound quanta. For this case Landau has shown that the absorption coefficient is proportional to $\omega^2$.

To summarize, we have generalized Landau's theory of Fermi liquids by taking into account the diffraction of de Broglie waves. To this end we used the quantum kinetic equation derived by us in recent paper \cite{tsin}. It should be noted that our kinetic equation is considerably reacher than the Landau kinetic equation. There is an additional physical feature included here, namely the Madelung term is incorporated due to the diffraction of de Broglie waves. This term is responsible for the excitation of zero sound even in an ideal Fermi gas. Thus we have generalized the Landau dispersion relation of zero sound including the quantum correction and found a new mode due to the quantum effect. We also disclosed a new branch of frequency spectrum due to the weak interaction.
Landau's theory of Fermi liquids is a vital theory of both theoretical and practical use. This theory has proven to be crucial for our understanding of a broad range of materials. Important examples of where this theory has been successfully applied are most notably electrons in most metals and Liquid $H_e^3$. The theory has found further application in nuclear and neutron star matter, superfluid $H_e^3$, and contemporary problems in superconductivity. Hence, the results of the present paper may be of substantial interest in connection with the above applications, as well as our theory is of great general theoretical interest.

\end{document}